 \documentclass[final,5p,times,twocolumn,authoryear]{elsarticle}

\IfFileExists{aas_macros.sty}{\usepackage{aas_macros}}{%

}

\usepackage{graphicx}
\usepackage{txfonts}
\usepackage{amssymb}
\usepackage{lipsum}
\usepackage{url} 
\usepackage{natbib}
\usepackage{amsmath}
\usepackage{caption} 
\usepackage{listings}
\usepackage{xcolor}

\usepackage[colorlinks=true,
linkcolor=black,     
citecolor=ForestGreen,
urlcolor=RoyalBlue,   
breaklinks=true]{hyperref}  

\lstdefinelanguage{json}{
    basicstyle=\ttfamily\small,
    numbers=left,
    numberstyle=\scriptsize\color{gray},
    stepnumber=1,
    numbersep=5pt,
    showstringspaces=false,
    breaklines=true,
    frame=single,
    backgroundcolor=\color{gray!10},
    literate=
     *{0}{{{\color{blue}0}}}{1}
      {1}{{{\color{blue}1}}}{1}
      {2}{{{\color{blue}2}}}{1}
      {3}{{{\color{blue}3}}}{1}
      {4}{{{\color{blue}4}}}{1}
      {5}{{{\color{blue}5}}}{1}
      {6}{{{\color{blue}6}}}{1}
      {7}{{{\color{blue}7}}}{1}
      {8}{{{\color{blue}8}}}{1}
      {9}{{{\color{blue}9}}}{1}
      {:}{{{\color{red}{:}}}}{1}
      {,}{{{\color{red}{,}}}}{1}
      {\{}{{{\color{black}{\{}}}}{1}
      {\}}{{{\color{black}{\}}}}}{1}
      {[}{{{\color{black}{[}}}}{1}
      {]}{{{\color{black}{]}}}}{1},
}
\DeclareCaptionLabelFormat{snippet}{Snippet #2} 
\newcommand{\citepnp}[1]{\citeauthor{#1}, \citeyear{#1}}

\journal{Astronomy $\&$ Computing}

\begin{document}
\begin{frontmatter}

\title{Encapsulating Textual Contents into a MOC data Structure for Advanced Applications}

\author[a]{Giuseppe Greco}
\author[b]{Thomas Boch}
\author[b]{Pierre Fernique}
\author[b]{Manon Marchand}
\author[b]{Mark Allen}
\author[b]{Francois‑Xavier Pineau}
\author[b]{Matthieu Baumann}
\author[c]{Marco Molinaro}
\author[d,e]{Roberto De Pietri}
\author[f,g]{Marica Branchesi}
\author[h,i]{Steven Schramm}
\author[j]{Gergely Dálya}
\author[k]{Elahe Khalouei}
\author[l]{Barbara Patricelli}
\author[m]{Giulia Stratta}

\address[a]{INFN Sezione di Perugia, via A. Pascoli, 23c, I-06123 Perugia, Italy}

\address[b]{Université de Strasbourg, CNRS, Observatoire astronomique de Strasbourg, UMR 7550, 67000 Strasbourg, France}

\address[c]{INAF - Osservatorio Astronomico di Trieste, 34143 Trieste, Italy}

\address[d]{INFN, Gruppo Collegato di Parma, Parco Area delle Scienze, 43124 Parma, Italy}

\address[e]{Università di Parma, Parco Area delle Scienze, 43124 Parma, Italy}

\address[f]{Gran Sasso Science Institute (GSSI), L'Aquila, Italy}

\address[g]{INFN - Laboratori Nazionali del Gran Sasso, 67100 Assergi, Italy}

\address[h]{DPNC, Section de Physique, Université de Genève, Geneva, Switzerland}

\address[i]{Gravitational Wave Science Center (GWSC), Université de Genève, Geneva, Switzerland}

\address[j]{L2IT, Laboratoire des 2 Infinis - Toulouse, Université de Toulouse, CNRS/IN2P3, UPS, F-31062 Toulouse Cedex 9, France}

\address[k]{Astronomy Research Center, Research Institute of Basic Sciences, Seoul National University, 08826 Seoul, South Korea}

\address[l]{Physics Department, University of Pisa, Largo B. Pontecorvo 3, 56127 Pisa, Italy}

\address[m]{INAF - OAS Bologna, Via P. Gobetti 93/3, 40129 Bologna (BO), Italy}

\begin{abstract}
\textit{Context.}
{
The Multi-Order Coverage map (MOC) is a widely adopted standard promoted by the International Virtual Observatory Alliance (IVOA) to support data sharing and interoperability within the Virtual Observatory (VO) ecosystem. This hierarchical data structure efficiently encodes and visualizes irregularly shaped regions of the sky, enabling applications such as cross-matching large astronomical catalogs, visualizing multi-wavelength and multi-messenger surveys, and facilitating collaborative research through seamless interoperability in big-data-driven exploration. }

\textit{Aims.}
{This study aims to explore potential enhancements to the MOC data structure by encapsulating textual descriptions and semantic embeddings into sky regions. 
Specifically, we introduce "Textual MOCs", in which textual content is encapsulated, and "Semantic MOCs" that transform textual content into semantic embeddings. These enhancements are designed to enable advanced operations such as similarity searches and complex queries and to integrate with generative artificial intelligence (GenAI) tools to improve context-aware interactions and response accuracy in astronomical data analysis, and support agent-based applications.}

\textit{Method. }
{We experimented with Textual MOCs by annotating detailed descriptions directly into the MOC sky regions, enriching the maps with contextual information suitable for interactive learning tools. For Semantic MOCs, we converted the textual content into semantic embeddings, numerical representations capturing textual meanings in multidimensional spaces, and stored them in high-dimensional vector databases optimized for efficient retrieval. 
}

\textit{Results.}
{The implementation of Textual MOCs enhances user engagement by providing meaningful descriptions within sky regions, facilitating the development of effective game-based learning. Semantic MOCs enable sophisticated query capabilities, such as similarity-based searches and context-aware data retrieval, enhancing astronomical data analyses. 
Integration with multimodal generative AI systems allows for more accurate and contextually relevant interactions 
supporting both spatial, semantic and  visual operations for advancing astronomical data analysis capabilities. Through straightforward examples, we discuss the fundamentals of this new experimental implementation.}

\end{abstract}

\end{frontmatter}

\section*{Introduction}

The International Virtual Observatory Alliance (IVOA)\footnote{\url{https://ivoa.net/}} is an international organization that develops and promotes open technical standards to enable seamless interoperability of astronomical data and tools within the Virtual Observatory (VO, \citepnp{ivoa}). Founded in 2002, IVOA is a free and open consortium that brings together scientific communities, research institutions, and software developers. It serves as a collaborative platform where members debate and define guidelines for data management, exchange, and analysis, ensuring that astronomical data is accessible and usable across different systems and platforms \citep{vo}. 

The Multi-Order Coverage Map (MOC) is a data structure recommended by the IVOA organization that provides a standardized method for mapping and representing sky regions based on the HEALPix sky tessellation \citep{gorski05}, promoting interoperability across astronomical datasets \citep{vo2}.

MOCs are highly efficient for representing complex sky regions at multiple levels of resolution, enabling rapid logical operations, seamless data processing, and fast querying capabilities. These advantages are further supported by IVOA data providers, who facilitate MOC-based queries, enabling swift access and retrieval of relevant astronomical data \citep{fernique2014, fernique2018}. As a common framework for mapping and querying astronomical datasets, MOC acts as a technical standard within the IVOA ecosystem, enhancing the FAIR (Findability, Accessibility, Interoperability, and Reusability)  principles, which are crucial for realizing the vision of a fully integrated and collaborative Virtual Observatory \citep{fair}.

Due to these properties, MOCs have been widely adopted in various fields of astronomical research \citep{grayson2023, carvajal21, smareglia19, smith17}, proving particularly significant in the emerging field of multi-messenger astronomy, especially for handling gravitational-wave sky localizations (Greco et al. \citeyear{greco2019}, \citeyear{greco2022a}; \citeyear{greco2022b}, \citepnp{singer2022}, \citepnp{coughlin2023}, \citepnp{saguaro2024}). Furthermore, MOCs are extensively utilized in interactive data visualization tools, such as Aladin Lite \citep{aladin_lite} and Aladin Desktop \citep{aladin20}, and have been integrated into online platforms like GWOPS \citep{gwops2020}, ESASky \citep{esasky}, GLADEnet \citep{gladenet}, 
Treasure Maps \citep{treasure_map}, GWsky\footnote{\url{https://virgo.pg.infn.it/maps/}}, and Fink\footnote{\url{https://fink-portal.org/gw}} \citep{fink}, among others. 

These interactive visualizations are enhanced by the ability to overlay a wide range of images or catalogs using the IVOA HiPS (Hierarchical Progressive Surveys) standard. HiPS is a widely used format in astronomy for representing and visualizing large datasets, such as images, catalogs and three-dimensional data cubes. HiPS technology divides spatial data into hierarchical tiles with progressive resolution, allowing efficient zooming and exploration of finer details as needed  \citep{hips}. 

MOC and HiPS are designed to be highly interoperable with each other. This interoperability allows users to seamlessly integrate MOC maps with HiPS data within visualization tools. Users can display MOCs alongside HiPS layers and customize graphic properties (e.g., transparency, outlines, color), enabling a more personalized and interactive exploration of astronomical data and facilitating skymap comparisons \citep{greco2022b}.

In this article, we report on our experimentation with a possible extension of the MOC standard to incorporate textual information, enabling simultaneous spatial and semantic operations. This new data structure, called "Textual MOCs," allows for the association of detailed descriptions with spatial data, enriching the context of astronomical information. This work is motivated by the possibility that these Textual MOCs can be transformed into "Semantic MOCs" by converting the textual content into semantic embeddings—numerical vectors that encapsulate the meaning of the text within high-dimensional spaces \citep{Word2Vec, glove, bert}. This transformation process unlocks significant potential for training generative AI models (GenAI), including large language models (LLM)-based systems \citep{goodfellow2014generative, brown2020language, lamda, opt}.
Semantic MOCs have the potential to significantly enhance query comprehension enabling advanced applications, such as
retrieval-augmented frameworks \citep{lewis2020retrieval, guu2020realm, karpukhin2020dense}, multi-agent orchestration \citep{multiagent}, and multimodal approaches \citep{multimodal}.
These references represent just a few examples from the extensive literature on the topic.

Each section in which the article is structured is complemented by a Jupyter notebook \citep{jupyter} with sample code, available for download to facilitate practical exploration and implementation of these techniques\footnote{https://github.com/ggreco77/TextualMOC}.

For large-scale operational applications, we refer to more detailed studies to be presented in future work, as this article primarily establishes the foundational concepts and outlines the key ideas of this approach.
\section{From MOCs to Textual MOCs}
\label{sec:basic_methods}
In this section, we provide a brief overview of the basic concepts for creating MOC maps.
Over time, the MOC structure has evolved to incorporate both spatial and temporal dimensions, leading to the development of the Space-Time MOC (ST-MOC) \citep{moc20}. 
However, in this work, we focus exclusively on MOCs with spatial indexing and will generally refer to (Space) MOCs in all examples. 
The textual$\slash$semantic extension of ST-MOCs is not presented here for brevity; as future work, we plan to employ it in agent-based applications that verify the availability of temporal information  and enable joint spatio-temporal operations.

\subsection{MOC data structure and serializations}
\label{subsec:mocstructure}
The MOC data structure is based on the HEALPix  sky tessellation algorithm \citep{gorski05}. HEALPix partitions the sphere into equal-area pixels, making it highly suitable for representing astronomical data on the celestial sphere.
In this algorithm, the sky is divided hierarchically into equal-area pixels, starting from twelve base pixels at the coarsest resolution level. Each subsequent level of resolution, indicated by the HEALPix parameter \textit{order}, divides each pixel into four smaller ones. 

Each MOC cell is defined by two properties:

\begin{itemize}
    \item \textbf{Hierarchy Level} (\textit{order}): Represents the level of refinement or resolution of the tessellation.
    \item \textbf{Pixel Index} (\textit{npix}): Uniquely identifies a pixel on the sky at a given resolution.
\end{itemize}

The finest level of refinement in the MOC hierarchy is determined by the \textit{order} parameter, allowing for an efficient representation of arbitrarily shaped regions of the sky with varying levels of detail.
The HEALPix orders typically range from 0 to 29, allowing for a mean cell resolution from 58.63 degrees down to 393.2 $\mu$as. 

To facilitate interoperability and ease of use across different platforms and applications, Multi-Order Coverage (MOC) supports multiple serialization formats. These formats enable the encoding and decoding of MOC data structures for storage, transmission, and visualization purposes. Two serialization formats are supported: $(i)$ FITS (Flexible Image Transport System) and $(ii)$ JSON (JavaScript Object Notation), as defined in the IVOA Recommendation \textit{MOC: Multi-Order Coverage map
Version 2.0 (2022-07-27)} \citep{fernique22}. Although JSON is not a normative serialization format in the MOC standard, it is widely used in practice and has already been employed in applications such as the tutorial by Greco $\&$ Marchand to compute the visibility of a MOC from a ground-based observatory as a function of airmass\footnote{\url{https://cds-astro.github.io/tutorials/7_Multi-messenger_astronomy__Planning_observations_rapidly_with_MOCs.html}}. In this work, we adopt JSON serialization when extensions are required, with the intention of extending our approach to the other serialization formats in future developments.

To store a MOC in a FITS file, each cell is converted into a single integer:

\begin{equation}
\text{uniq} = 4 \times 4^{\textit{order}} + \textit{npix}
\end{equation}

A MOC, when serialized in JSON format, follows the following syntax:

\begin{lstlisting}[language=json, numbers=none, caption={JSON structure of a Space MOC}, label={fig:moc}, mathescape=true]
{
  "order$_n$": [npix$_i$ | where npix$_i$ uniquely defined integers],
  // ... continue for other orders ...
}
\end{lstlisting}

\noindent In this JSON structure, shown in Snippet \ref{fig:moc}, each \colorbox{gray!20}{order$_n$} represents a different resolution level, with \colorbox{gray!20}{npix$_i$} indicating the pixel indices covering specific sky regions at that order. 

\subsection{Textual MOC data structure}
\label{subsec:mocstructure_2}
For the creation of Textual MOCs, we decide to use the JSON serialization format.
JSON  is a lightweight, human-readable data interchange format that is language-independent and features a simple syntax, making it ideal for transmitting data over the web. Its self-describing nature and minimal structure make JSON an excellent choice for efficiently encoding and sharing information. 

Snippet \ref{fig:textualMOC} illustrates the JSON structure of a Textual MOC. In addition to the first two entries (\colorbox{gray!20}{order$_n$} and \colorbox{gray!20}{npix$_i$}) discussed in Subsection \ref{subsec:mocstructure}, text-based MOCs contain the following new elements:
\begin{description}
    \item[\colorbox{gray!20}{text}] Provides a textual description of the content and scientific relevance of the celestial region or dataset.
    \item[\colorbox{gray!20}{metadata}] Contains additional information such as the author, creation date, and other key-value pairs for metadata enrichment.
    \item[\colorbox{gray!20}{annotated\_cells}] Maps specific MOC orders and pixels to user-defined annotations, allowing labels to be attached to individual sky cells.
\end{description}

While the \colorbox{gray!20}{text} key provides a global annotation across the entire MOC coverage, the \colorbox{gray!20}{annotated\_cells} key facilitates the assignment of labels to specific MOC cells. 
In multi-agent workflows, the system can be instructed to identify particular objects referenced in the text, retrieve their coordinates from VO services such as SIMBAD, and annotate the corresponding cells. In addition, the MOC resolution can be increased to achieve more detailed coverage when required.

A possible use case can be considered when, within the localization region of a gravitational wave event, a set of candidate objects is identified. These candidates can be inserted into the keyword \colorbox{gray!20}{annotated\_cells} to enable specific searches in databases such as Vera Rubin, or to check for coincident high-energy events reported by satellite observatories. This type of annotation enables automated searches, for example on kilonova events or short gamma-ray bursts, because the textual information embedded in a MOC allows the search to be properly contextualized within CBC (Compact Binary Coalescence) gravitational-wave events.

\begin{lstlisting}[language=json, numbers=none, firstnumber=1, caption={JSON structure of a Textual MOC with text, metadata, and cell annotations}, label={fig:textualMOC}, mathescape=true]
{
  "order$_n$": [npix$_i$ | where npix$_i$ uniquely defined integers],
  // ... continue for other orders ...
  "text": "Your textual description here",
  "metadata": {
    "key1": "value1",
    "key2": "value2",
    // ... continue metadata entries ...
  },
  "annotated_cells": {
    "order$_n$": {
      "npix$_i$": "annotation$_i$"
      // ... continue for other pixels ...
    }
    // ... continue for other orders ...
  }
}
\end{lstlisting}

In the accompanying tutorial, we introduce basic Python functions to encapsulate textual information within MOCs, demonstrating how to directly include this data in the JSON representation. These functions utilize the \texttt{mocpy} \citep{mocpy}, \texttt{matplotlib} \citep{matplotlib}, and \texttt{astropy} \citep{astroquery} libraries, along with Python's native \texttt{json} library\footnote{\url{https://docs.python.org/3.12/library/json.html}}.

\section{The Role of MocServer in Creating Textual MOCs}
\label{subsec:mocserver}
In this section, we recall the main features of MOCserver, a server specifically dedicated to MOCs. In our context, this is particularly valuable as it allows us to create textual MOCs with metadata curated by archival scientists, ensuring both accuracy and reliability. We present an example application showcasing both the embedded textual content within the MOC and its spatial component simultaneously. 
MOCs can describe the spatial coverage of any dataset; when generated from images, the corresponding images in HiPS format can be retrieved. To render the spatial component, we use ipyaladin, a Jupyter/IPython widget that enables interactive visualization of HiPS and MOCs directly within notebook environments \citep{ipyladin}.

\subsection{MocServer}
\label{subsec:mocserver}
A dedicated directory service, MOCserver\footnote{\url{https://alasky.cds.unistra.fr/MocServer/query}}, is specifically designed for the storage, curation, and preservation of MOC data. MOCserver ensures that MOC datasets are securely maintained, while also facilitating easy access and retrieval for further analysis and use. 

The MocServer is developed by the Strasbourg Astronomical Data Centre (CDS) in 2015, designed to efficiently manage collections of spatial and temporal data encoding with MOC data structures. Its primary objective is to quickly provide a list of relevant astronomical resources for a given region of the sky, with response times in milliseconds. MocServer stores dataset names, each linked with a MOC (Multi-Order Coverage) spatial coverage map and detailed metadata.

This service is widely utilized by various tools, such as Aladin Desktop\footnote{\url{https://aladin.cds.unistra.fr/AladinDesktop/}}, Aladin Lite\footnote{\url{https://aladin.cds.unistra.fr/AladinLite/}},  and the CDS portal\footnote{\url{http://cdsportal.u-strasbg.fr/}}, which leverage MocServer's capabilities to provide quick access to a wealth of astronomical data. As of 2023, MocServer contains more than 32,000 entries, underscoring its significance within the astronomical community.

This service can be accessed remotely through an \texttt{HTTP} API. 
Additionally, a Python module is available on GitHub as part of the \texttt{astroquery} \citep{astroquery}, providing a convenient interface for integrating this functionality into Python-based workflows. 

The result of such a query can go beyond an identifier-only list, extending to the properties themselves and effectively offering a "metadata directory service." The use of MocServer keys is based on the IVOA model and relies extensively on its vocabulary. 
A comprehensive list of metadata keywords managed by the MocServer, along with their occurrence counts and an example of the associated value for each keyword, can be obtained through the API. This is achieved by appending the parameter 
\colorbox{gray!20}{get=example}
to the MocServer URL, as shown in the following example: \url{https://alasky.cds.unistra.fr/MocServer/query?get=example}.

Furthermore, we emphasize that a lightweight and flexible local infrastructure for MOC-based coverage queries can be built with texttt{MOC-Set}\footnote{\url{https://pypi.org/project/moc-set/}}, which stores only the MOC regions without the associated metadata provided by the MOCServer. This limitation can be readily addressed by storing metadata separately (e.g. SQLite, PostgreSQL, MongoDB, or other lightweight database systems) and retrieving them via the list of MOC identifiers returned by MOC-Set queries. A detailed analysis of these technologies lies beyond the scope of this work; here we simply demonstrate the broad support of the MOC standard within the VO ecosystem and show how the proposed extension can be seamlessly integrated into it.

\subsection{Textual MOC powered by MOCserver}
\label{subsec:mocserver}
We have experimented with the MocServer as a web tool for creating Textual MOCs because it provides access to a wide range of curated datasets with reliable references and detailed metadata. This access allows researchers to incorporate authoritative data into their MOCs, enhancing them with scientifically accurate and comprehensive information. 

\begin{figure*}
	\centering 
	\includegraphics[width=1\textwidth,height=0.5\textwidth]{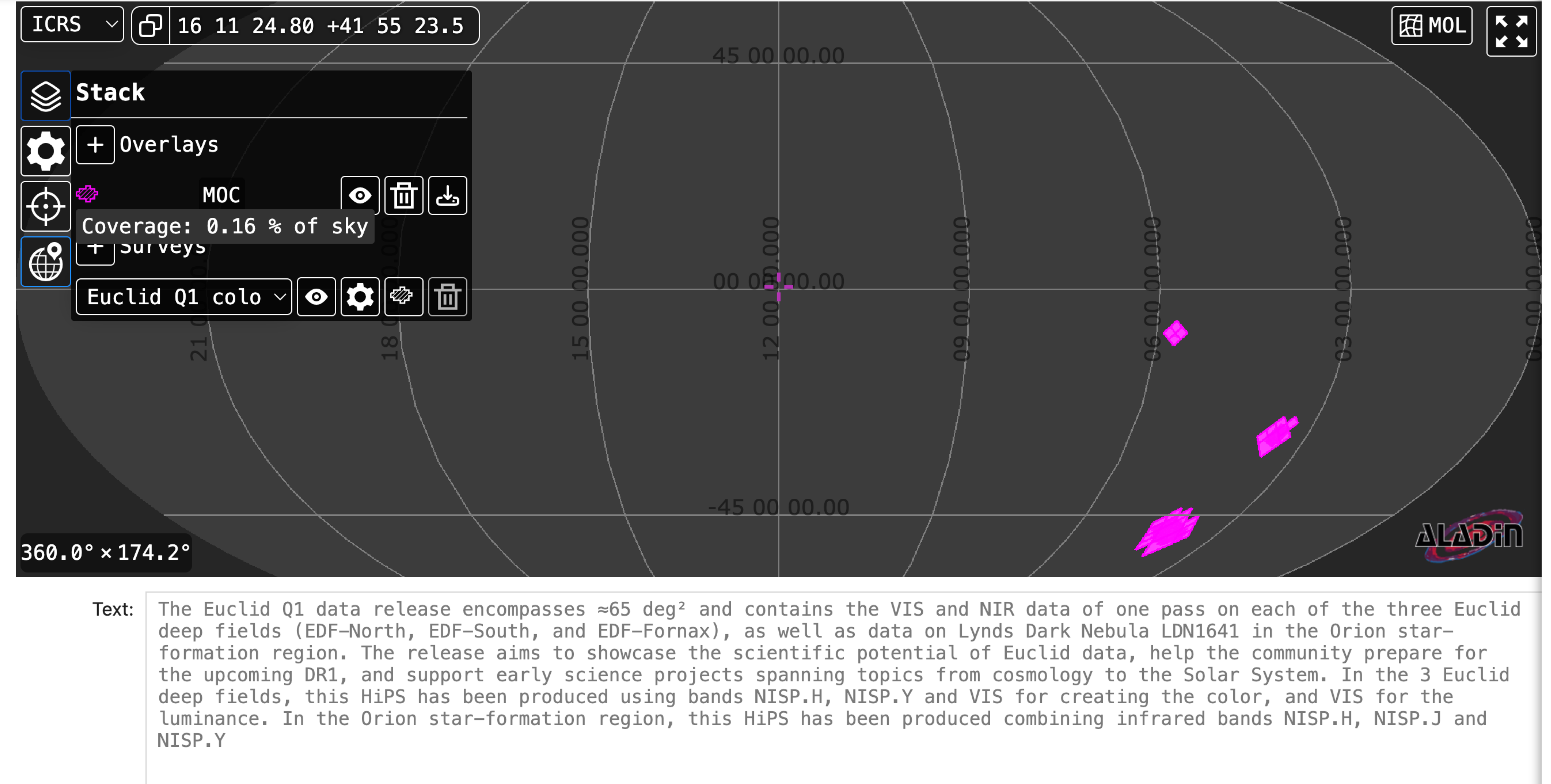}	
	\caption{
 An illustration showing how a Multi-Order Coverage (MOC) map can be enriched with textual content retrieved from MOCServer, allowing the simultaneous visualization of both spatial and textual components. The MOC for the dataset \texttt{CDS/P/Euclid/Q1/color} (Euclid Quick Data Release 1) from the Euclid space survey mission is displayed, with metadata (specifically the \colorbox{gray!20}{obs\_description} field) used to generate a textual MOC. The interactive visualization is provided by ipyaladin, a Jupyter/IPython widget, showing both MOCs and their corresponding image datasets in HiPS format. 
 } 
	\label{fig_1}%
\end{figure*}

Figure \ref{fig_1} illustrates how a Multi-Order Coverage (MOC) map can be enriched with textual content retrieved from the MOCserver, enabling simultaneous visualization of both textual and spatial components. 
 
As an illustrative example, we retrieve the MOC for the dataset \texttt{CDS/P/Euclid/Q1/color} via MOCServer, which provides both the spatial coverage map (MOC) and its associated metadata. This dataset originates from the Euclid mission provided by the ESA (European Space Agency). In this case, the metadata field \colorbox{gray!20}{obs\_description} is utilized to create the textual MOC. 

\section{Astronomical Platforms for Textual MOC Generation}
\label{subsec:gw_alert}
Other astronomical platforms can be utilized for the creation of Textual MOCs, such as the General Coordinates Network (GCN)\footnote{\url{https://gcn.nasa.gov/}}, the Astronomer’s Telegram (ATel)\footnote{\url{https://www.astronomerstelegram.org/}}, the Transient Name Server (TNS) \citep{tns}, the Astrophysical Multimessenger Observatory Network (AMON) \citep{amon} and follow-up pointings in the Gravitational Wave Treasure Map \citep{treasure_map}.

As a user case, consider displaying 90\% of a credible region of a gravitational-wave sky localization simultaneously with an associated alerts of the LIGO, Virgo and KAGRA collaborations (LVK)\footnote{\url{https://emfollow.docs.ligo.org/userguide/}} via The General Coordinates Network (GCN). The GCN Circular text is retrieved through the site's functionality, which provides access to it in both text and JSON formats.
 The method for generating MOCs of sky localization area of gravitational-wave events is described in detail by \citep{greco2022b}. In this work, we extend that approach by integrating textual data with spatial regions through Textual MOCs, combining the corresponding GCN Circular(s) with the corresponding sky localization MOC map.

\section{Enhancing Data Exploration with Textual MOCs and Aladin Lite}
\label{sec:aladin_application}
In Section 2, we have explored the use of ipyaladin (a Jupyter/IPython widget) to simultaneously display the spatial and textual content of MOCs. 
Here, we utilize Aladin Lite in combination with Textual MOCs to develop web applications that enhance interactivity. 

We use version 3 of Aladin Lite, which introduces a major overhaul of the platform by leveraging GPU acceleration through WebGL enhancing performance and interactivity \citep{aladinv3}.

\subsection{Contextualization and Interactivity}
\label{subsec:game}
\begin{figure*}
	\centering 
	\includegraphics[width=0.8\textwidth,height=0.5\textwidth]{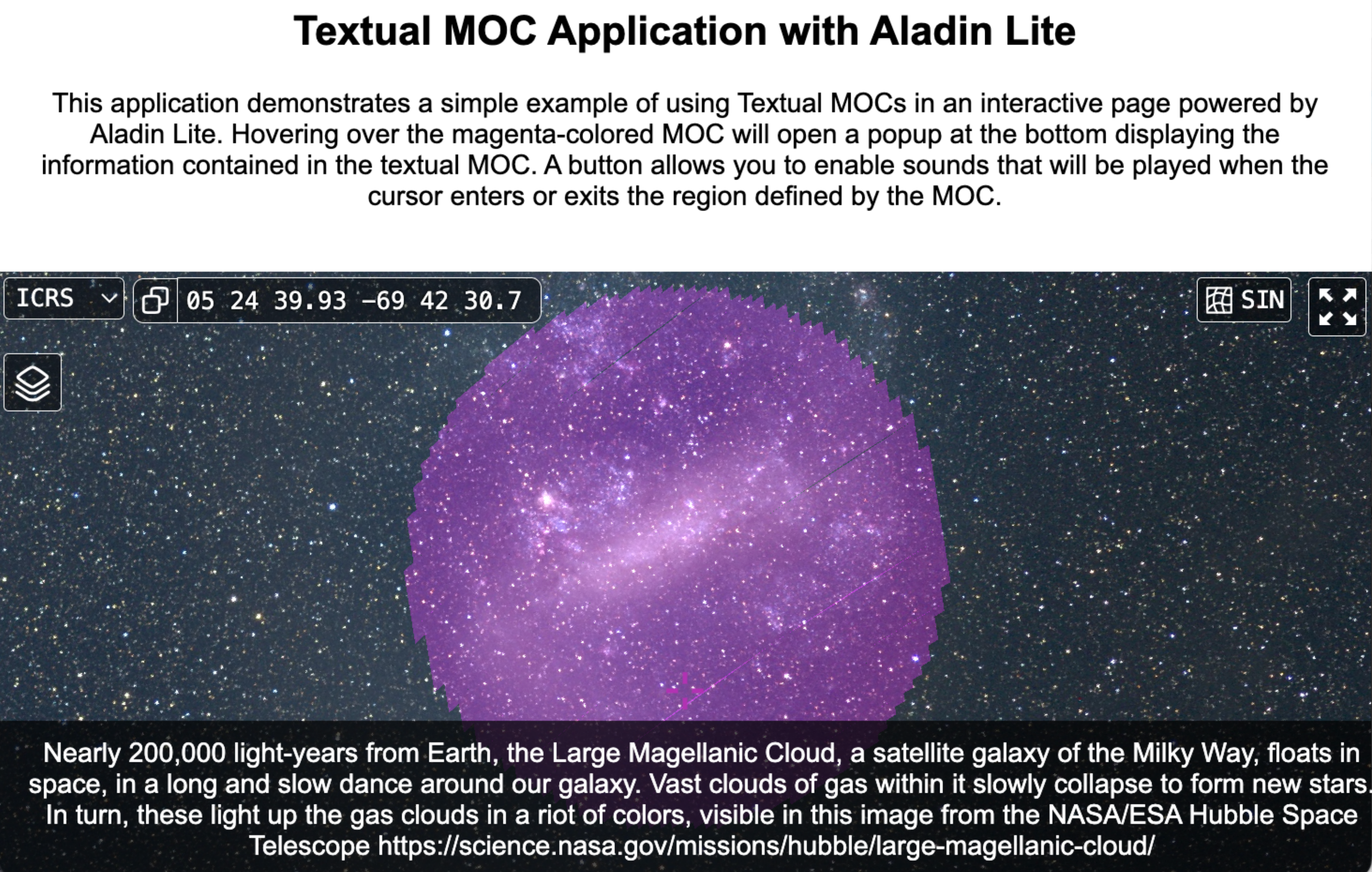}
	\caption{An illustrative example demonstrating the integration of the Textual MOC data structure into a web page using Aladin Lite. The figure illustrates the Aladin Lite viewer with the MOC map, an informative popup activated by the cursor, and a menu for managing sound effects. Interactive demo: \url{https://ggreco77.github.io/TextualMOC/AladinGame/aladin_game_d1/} } 
	\label{fig_mom0}%
\end{figure*}

As a minimal example, we show how to integrate and utilize Textual MOC data structure within a web page using Aladin Lite. As shown in Figure \ref{fig_mom0}, the HTML structure is designed to include essential elements such as a container for the Aladin Lite viewer, where the MOC map is displayed, and an informative pop-up that appears when the cursor hovers over a MOC region. The information depicted in the pop-up is extracted from the textual component of the MOC, enabling simultaneous access to both spatial and descriptive data.

Additionally, it manages sound effects, providing auditory feedback when the mouse enters or exits a MOC region, thereby enhancing the application's overall interactivity.
Users can then interact with the Aladin Lite viewer, explore the map, view contextual information in the pop-up, and control sound effects via the dedicated menu. This implementation effectively combines visual, textual, and auditory elements, creating an immersive and interactive experience for exploring data within a Textual MOC structure.

Both MOCs and HiPS in Aladin Lite utilize level-of-detail (LOD) techniques, dynamically adjusting the resolution based on zoom level. Lower resolution is used for large fields of view to ensure smooth interaction, while higher resolution is progressively loaded as users zoom in, optimizing both performance and data visualization. This flexibility supports efficient comparison of datasets and simultaneous analysis of multiple layers.

\subsection{Learning Game and Public Engagement}
\label{subsec:game}

\begin{figure}
 \includegraphics[width=0.49\textwidth,height=0.55\textwidth]{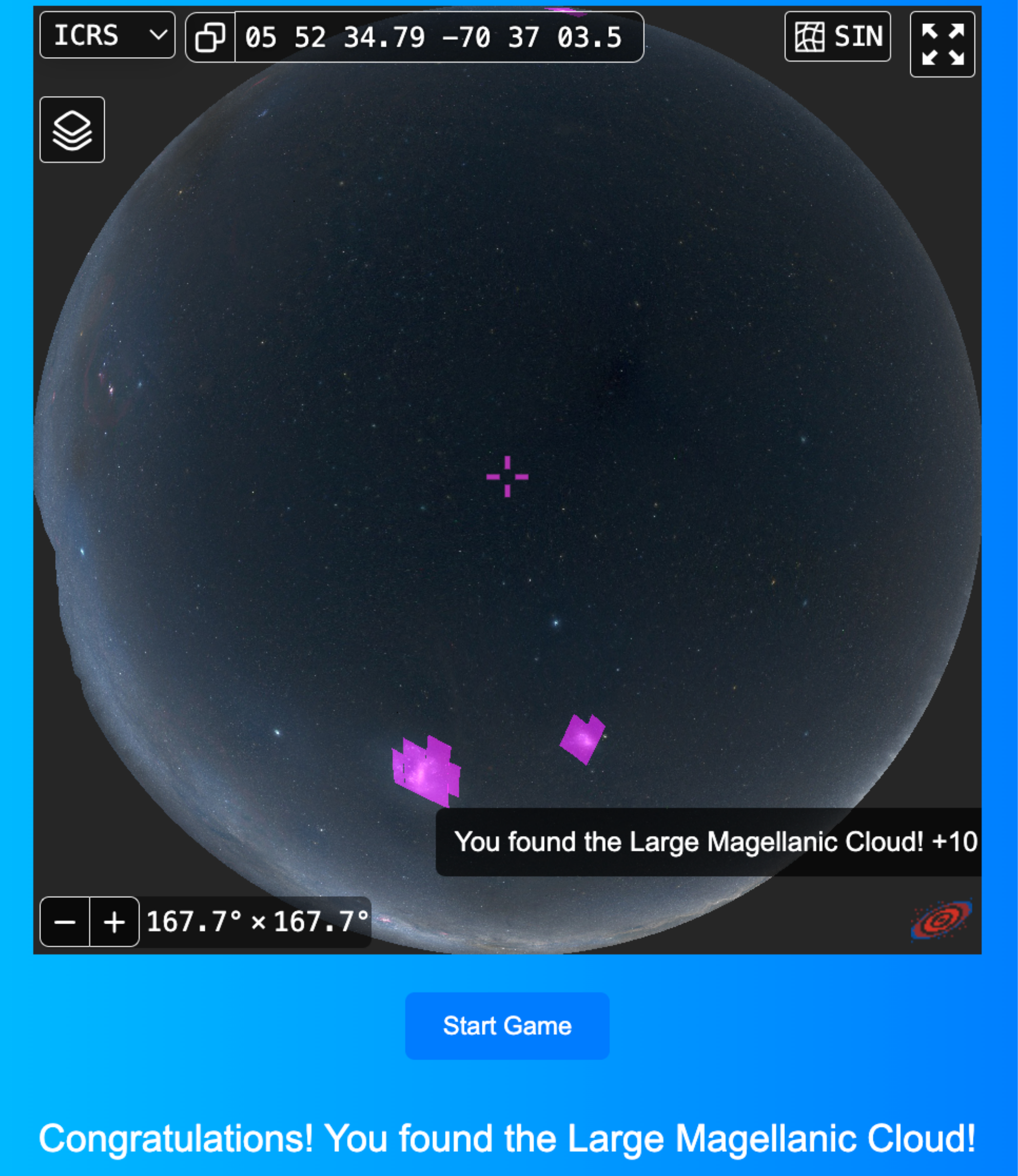}
	\caption{An interactive game using Textual MOCs within Aladin Lite helps users locate the Large Magellanic Cloud (LMC). The magenta-colored MOC marks the position of the LMC, and informational popups provide feedback as the user interacts with the map. Auditory cues enhance the user experience by signaling correct or incorrect selections.
    Interactive demo: \url{https://ggreco77.github.io/TextualMOC/AladinGame/aladin_game_d2/}} 
	\label{fig_mom1}%
\end{figure}

Building upon the previous example, we developed an interactive game demonstrator that guides users in locating the Large Magellanic Cloud (LMC) within the sky.
The game challenges users to navigate through a sky map where multiple regions are covered by semitransparent MOCs. These MOCs represent various areas of the sky, and the user's objective is to identify the region corresponding to the LMC. When the cursor hovers over the magenta-colored MOC that overlays the LMC, a pop-up displays a message confirming the discovery, and the user's score is updated. Auditory feedback is also incorporated to signal correct or incorrect selections, enhancing the interactive aspect of the game,  see Figure \ref{fig_mom1}.

This implementation showcases the effective use of Textual MOCs in an interactive learning setting. Users can visually explore the sky, guided by real-time feedback provided by the MOCs. The combination of spatial and textual data not only promotes an engaging activity but also serves as a practical game-based learning, highlighting the application of astronomical data in an interactive format. The examples can be reproduced using the tutorial associated with the article.

\section{MOCs in the Generative AI Stack: From Embeddings to Multimodal RAG and Agents}
\label{sec:powered_ai}

Here we discuss how these concise textual annotations within MOCs can now be utilized by GenAI models. To enable this, the text must be converted into numerical vectors, as the processing method in GenAI relies on embeddings—vectors (lists) of floating-point numbers that capture semantic information and meaning.

We demonstrate how the Textual MOCs can be utilized within multimodal Generative AI (GenAI) frameworks, enabling simultaneous semantic, spatial and image-based visual recognition  tasks. We show how to transform the textual component into semantic embeddings, store them in vector databases, and perform basic operations such as retrieval and classification.
The HiPS standard, adopted by the IVOA for storing and visualizing astronomical images, will be used to integrate image URLs into semantic MOCs. This integration will be implemented through the \texttt{hips2fits}\footnote{\url{https://alasky.cds.unistra.fr/hips-image-services/hips2fits}} server, allowing seamless access to corresponding images.
With this approach, multimodal models—especially those optimized for visual recognition, image reasoning, caption generation, and question answering on images—can be effectively utilized.

\subsection{Semantic MOC}
\label{semanticMoc}

Textual MOCs are converted into semantic embeddings, which are compact numerical representations that encapsulate the meaning of the text within a multidimensional geometric space \citep{Word2Vec, glove, bert}. 
Various approaches exist for generating semantic embeddings, enabling the numerical representation of text, images, audio, or structured data. The choice of method depends on the data type, the model used, and the specific application.  

To ensure reproducibility in the generation of Semantic MOCs, we provide a Python function that converts the text field of a MOC into a numerical embedding and stores it within the same data structure. 
The function employs the \texttt{LangChain} framework with the \texttt{OllamaEmbeddings} integration \footnote{\url{https://python.langchain.com/docs/integrations/text_embedding/ollama/}}, allowing researchers to select among different embedding models depending on the accuracy and computational cost requirements.
The function \texttt{embedding\_from\_custom\_text}  reads the \colorbox{gray!20}{text} key from the MOC JSON object, generates the embedding vector using the specified model (default: \texttt{nomic-embed-text}), and stores both the resulting vector and the model name under the dedicated keys (\texttt{embedding}, \texttt{embedding\_model}). This guarantees a straightforward and traceable conversion from Textual to Semantic MOC, since the applied model is explicitly documented in the assigned keys.
Different embedding models can be employed depending on the desired trade-off between accuracy and efficiency (for example, \texttt{all-MiniLM-L6-v2}\footnote{\url{https://huggingface.co/sentence-transformers/all-MiniLM-L6-v2}}, \texttt{nomic-embed-text}\footnote{\url{https://huggingface.co/nomic-ai/nomic-embed-text-v1}}, \texttt{mistral-embed models}\footnote{\url{https://docs.mistral.ai/capabilities/embeddings/}}
, among others),
which can be consulted on the providers’ official documentation for further details. 
This compact representation enables efficient storage and indexing of embeddings in modern vector databases.

\begin{lstlisting}[language=json, numbers=none, firstnumber=1, caption={JSON structure of a Semantic MOC with text transformed into semantic embeddings.}, label={fig:semanticMOC}, mathescape=true]
{
  "order$_n$": [npix$_i$ | where npix$_i$ uniquely defined integers],
  // ... continue for other orders ...
  "text": "Your textual description here",
  "metadata": {
    "key1": "value1",
    "key2": "value2",
    // ... continue metadata entries ...
  },
  "annotated_cells": {
    "order$_n$": {
      "npix$_i$": "annotation$_i$"
      // ... continue for other pixels ...
    }
    // ... continue for other orders ...
  },
  "embedding": [text vectorization],
  "embedding_model": "selected embedding model"
  "image": "https://alasky.cds.unistra.fr/../..",
}
\end{lstlisting}

To specify semantic MOCs, we introduce three new entries in the serialization of textual MOCs, as illustrated in Snippet \ref{fig:semanticMOC}. 

The \colorbox{gray!20}{embedding} field contains the numerical vector representation of the text, generated through a chosen text vectorization technique, encoding the semantic meaning in a compact form suitable for computational analysis. 

The \colorbox{gray!20}{embedding\_model} entry specifies the embedding model used to generate the vector representation.
The choice of model determines the characteristics of the embedding and its applicability to different data types.

The \colorbox{gray!20}{image} key links to an astronomical image. In this scenario, the hips2fits\footnote{\url{https://alasky.cds.unistra.fr/hips-image-services/hips2fits}} server is used to retrieve astronomical images via direct URL parsing, thereby ensuring fast and consistent data access. 
The hips2fits service enables the dynamic generation of FITS image cutouts of arbitrary size, resolution, and field of view from a specified HiPS (Hierarchical Progressive Survey) dataset. 
Each image is properly accompanied by a comprehensive, source-specific bibliography.

In the next section, we show how these new fields enable the use of MOCs in generative AI models, not only text-based but also those with visual capabilities.

\subsection{Vector database, RAG, and Visual Models}

In the context of generative AI and LLMs, an increasingly widespread approach to querying document collections is the Retrieval-Augmented Generation (RAG) paradigm \citep{lewis2020retrieval,guu2020realm,karpukhin2020dense}. RAG combines embedding-based semantic retrieval with automatic text generation via advanced LLMs. In this framework, each document is transformed into a numerical vector—an embedding—that captures its semantic content using pre-trained models such as Sentence-BERT \citep{reimers2019sentence}. These vectors are stored in a vector database, which is optimized to quickly return the documents most similar to a query based on semantic similarity \citep{vectordatabase}.
Upon receiving a query, the system encodes it as an embedding, retrieves the most relevant documents, and supplies them as context to a generative model. This approach enables the generation of responses grounded in concrete sources, thereby reducing so-called “hallucinations”—unfounded assertions produced by the model \citep{shuster2021retrieval}. By employing specialized vector storage solutions, the pipeline remains scalable and can be refreshed in near real-time, supporting low-latency queries over extensive text corpora while ensuring content stays up to date.

In our approach, input documents are represented as Textual MOCs and processed through a conventional Retrieval-Augmented Generation (RAG) workflow. We enhance this pipeline with a dedicated spatial filter in the vector database
enabling hybrid queries that jointly consider semantic relevance and spatial location.

\begin{figure}[ht]
  \centering
  \includegraphics[width=\linewidth]{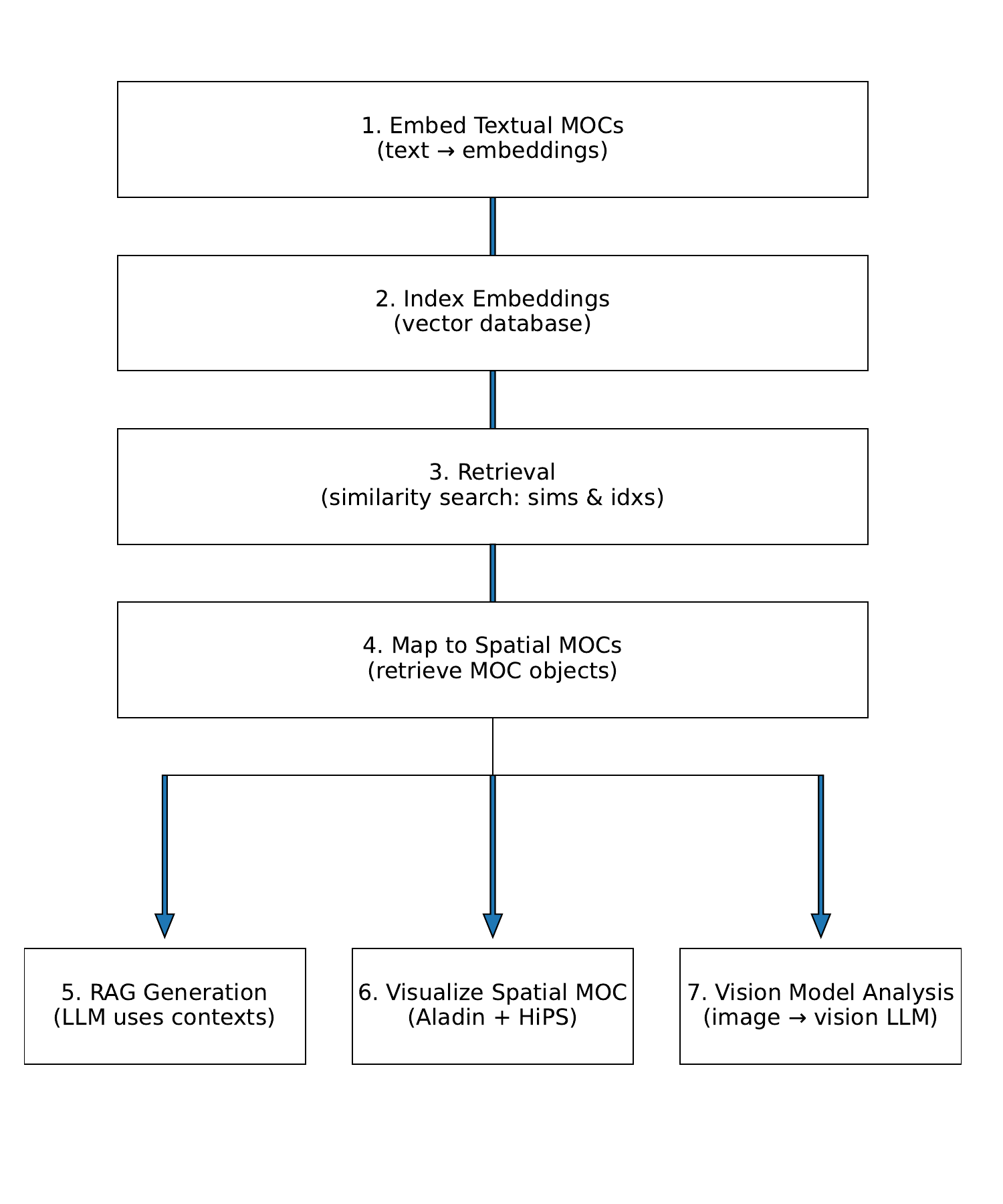}

  \caption{RAG Processing Framework with spatial MOC integration and vision analysis.  
  From a user query, (1) we embed textual MOCs into vector representations and (2) index them in a vector database.  
  In (3) a retrieval step returns a set of similarity scores (\(\text{sim}_k\)) and their corresponding record indices (\(\text{idx}_k\)), where each \(\text{sim}_k\) quantifies the relevance of the \(k\)-th MOC to the query and \(\text{idx}_k\) points to the original MOC entry.  
  These indices allow us to (4) fetch the matching spatial MOC objects, (5) generate context-grounded answers with a language model, (6) overlay the selected MOCs on all-sky imagery via Aladin Lite, and (7) stream the FITS image URLs (\texttt{image}) to a vision LLM for automated visual feature analysis.}
  \label{fig:RagMocPipeline}
\end{figure}

Specifically, from a user query, the system first encodes the Textual MOCs into embedding vectors (Step 1 - see Section \ref{semanticMoc}) and indexes them in a vector database optimized for similarity search (Step 2). During the retrieval step (Step 3), a set of similarity scores $\mathrm{sim}_k$ and their corresponding record indices $\mathrm{idx}_k$ is returned: each $\mathrm{sim}_k$ quantifies the relevance of the $k$-th textual MOC to the query, while $\mathrm{idx}_k$ points to its original entry in the database.
These indices then allow the system to fetch the associated spatial MOC objects (Step 4) and use them as context to generate grounded responses via a language model (Step 5). Concurrently, the selected MOCs are overlaid on an all-sky image using Aladin Lite (Step 6), providing an immediate visual representation of the queried region. Finally, the FITS image URLs (\texttt{image}) are streamed to a vision-enabled LLM (Step 7) for automated analysis of visual features. The complete pipeline—from textual query to sky map and visual analysis—is summarized in Figure~\ref{fig:RagMocPipeline}.

In the basic tutorial, we present a proof-of-concept system implementing a multimodal Retrieval-Augmented Generation (RAG) pipeline that combines spatial and semantic retrieval with visual enrichment via automated image analysis. We construct ten spatial Multi-Order Coverage (MOC) footprints centered on well-known galaxies; for each object we provide a $\sim$ 150-word textual summary and a link to an image obtained via \texttt{hips2fits}. The pipeline integrates lightweight local models for embeddings, text generation, and vision, supporting end-to-end retrieval and interpretation.
The system supports two complementary query modes. 
$(i)$ Semantic mode accepts natural-language queries (e.g., “Check whether the database contains the host galaxy of the gravitational-wave event GW170817.  If yes, provide a concise summary of the corresponding document.”). Object descriptions are indexed in a vector search index built from sentence embeddings; the RAG step selects the most relevant documents, and a local LLM produces the answer. To ensure transparency, the LLM output enforces a trailing USED: line that enumerates only the documents actually cited. The MOC footprints of the USED documents are overlaid in ipyaladin, and their corresponding images are downloaded and supplied to a vision model for morphological classification (spiral/elliptical/irregular) and detection of tidal features. 
$(ii)$ Positional mode accepts sky coordinates (RA, Dec) and tests containment in available MOCs using the mocpy library. Matching footprints are rendered in ipyaladin, and the associated images are analyzed by the same vision model.
The proof-of-concept is not intended as a comprehensive validation of embedding, vision, or textual models; rather, it demonstrates that lightweight, locally hosted models can be chained to deliver coherent, inspectable results (retrieval → spatial overlay → visual reasoning). Finally, the system can be designed to scale toward automatic result collection, alert dissemination, and real-time specialist chatbots, using an extension of a well-known and well-documented IVOA standard (MOC)—enriched with textual contents and image references—while preserving full operational compatibility.

\section{Conclusions and Future Outlook}
In this paper, we explore several potential applications of a widely adopted IVOA standard for encoding complex sky regions—namely, MOC.
Our primary goal is to investigate how this data structure can be efficiently integrated into modern generative AI architectures.
We introduce an experimental data structure, which we refer to as Textual MOCs when raw text annotations are applied directly, or Semantic MOCs when that text is converted into semantic embeddings via pre-trained models and encapsulated within the same MOC—making it more suitable for storage in vector databases. More specifically, 
the textual portion, encapsulated through specific keys within the spatial MOC, is then converted into semantic embeddings, which can be employed, for example, in RAG applications for the joint retrieval of MOC information, semantic data, and spatial coordinates. Moreover, Virtual Observatory servers (e.g., hips2fits) allow the generation of URLs for images from various astronomical surveys and, within the same workflow, direct these requests to vision models for further analysis. In this work, we implement JSON serialization of the MOC; future developments may consider employing FITS serialization by embedding the additional metadata within the header or by structuring ASCII tables with MOCs in one column and the corresponding metadata in additional columns.

 Textual/Semantic MOCs are currently
a practical tool for embedding textual information into spatial data. They have the potential to evolve into a formalized model aligned with the IVOA  proposed standards. When integrated into the broader IVOA Resource Metadata framework (VOResource) \citep{2025ivoa.spec.0416D}, Textual and Semantic MOCs can fully comply with FAIR principles (Findable, Accessible, Interoperable, Reusable), allowing tools and services to access and combine spatial and textual information seamlessly and efficiently.

We note that Textual/Semantic MOCs inherently produce structured outputs, making them ideal building blocks for agent-based systems. In this paradigm, each MOC can function as an autonomous software agent—encapsulating its embeddings, textual metadata, and behavior (e.g., relevance scoring, summary generation, visualization)—and responding to incoming queries or environmental triggers with discrete actions.

Looking ahead, we plan to conduct dedicated simulations of these data-driven agents in the context of third-generation interferometers (Einstein Telescope \citep{et}, Cosmic Explorer \citep{cosmic_explorer}) and space-based missions (e.g., LISA, THESEUS, ATHENA). In such scenarios, multi-messenger inputs—from neutrinos and GRBs to gravitational waves—must be efficiently orchestrated across varied time windows, with rapid, agent-triggered alerts for follow-up observations. By treating each MOC as a self-contained agent, we achieve a naturally distributed, scalable, and robust architecture capable of meeting the demands of next-generation multi-messenger astronomy.

\section*{Acknowledgements}
The authors acknowledge support from PNRR \textbf{ETIC} (IR0000004), “Einstein Telescope Infrastructure Consortium” — CUP I53C21000420006 (MISSIONE 4, COMPONENTE 2, INVESTIMENTO 3.1); from the Astrophysics Centre for Multi-messenger Studies in Europe (\textbf{ACME}), funded by the European Union’s Horizon Europe Research and Innovation Programme (Grant Agreement No. 10113192); from the \textbf{ESCAPE} project (European Science Cluster of Astronomy \& Particle Physics ESFRI Research Infrastructures), funded by the European Union’s Horizon 2020 research and innovation programme (Grant Agreement No. 824064);  and from the Italian National Institute for Nuclear Physics (INFN) and the Italian Space Agency (ASI) under the \textbf{ASI-INFN} agreement No. 2021-43-HH.0.

\bibliographystyle{elsarticle-harv} 
\bibliography{biblio}
\end{document}